\documentclass[%
 reprint, amsfonts,amssymb,amsmath,longbibliography,
superscriptaddress,
 amsmath,amssymb,
 aps,
prb,
]{revtex4-2}

\usepackage{graphicx}
\usepackage{dcolumn}
\usepackage{bm}
\usepackage[colorlinks=true,citecolor=blue,linkcolor=blue,urlcolor=blue]{hyperref}
\usepackage{color}
\usepackage{xspace}

\newcommand{\BZRO}{Ba$_{3}$ZnRu$_{2}$O$_{9}$\xspace}

\newcommand{\BMRO}{Ba$_{3}M$Ru$_{2}$O$_{9}$\xspace}
\newcommand{\BRRO}{Ba$_{3}R$Ru$_{2}$O$_{9}$\xspace}
\newcommand{\kbT}{k_{\rm B}T\xspace}

\begin{document}


\title{Magnetic ground state of the dimer-based hexagonal perovskite Ba$_{3}$ZnRu$_{2}$O$_{9}$}

\author{S.~Hayashida}
\email[]{s\_hayashida@cross.or.jp}
\altaffiliation[Present address: ]{Neutron Science and Technology Center, Comprehensive Research Organization for Science and Society (CROSS),
Tokai, Ibaraki 319-1106, Japan.}
\affiliation{Max-Planck-Institute for Solid State Research, Heisenbergstra$\beta$e 1, 70569 Stuttgart, Germany}
\author{H.~Gretarsson}
\affiliation{Max-Planck-Institute for Solid State Research, Heisenbergstra$\beta$e 1, 70569 Stuttgart, Germany}
\affiliation{PETRA III, Deutsches Elektronen-Synchrotron DESY, Notkestra$\beta$e 85, 22607 Hamburg, Germany}
\author{P.~Puphal}
\affiliation{Max-Planck-Institute for Solid State Research, Heisenbergstra$\beta$e 1, 70569 Stuttgart, Germany}
\author{M.~Isobe}
\affiliation{Max-Planck-Institute for Solid State Research, Heisenbergstra$\beta$e 1, 70569 Stuttgart, Germany}
\author{E.~Goering}
\affiliation{Max-Planck-Institute for Solid State Research, Heisenbergstra$\beta$e 1, 70569 Stuttgart, Germany}
\author{Y.~Matsumoto}
\affiliation{Max-Planck-Institute for Solid State Research, Heisenbergstra$\beta$e 1, 70569 Stuttgart, 
Germany}
\author{J.~Nuss}
\affiliation{Max-Planck-Institute for Solid State Research, Heisenbergstra$\beta$e 1, 70569 Stuttgart, Germany}
\author{H.~Takagi}
\affiliation{Max-Planck-Institute for Solid State Research, Heisenbergstra$\beta$e 1, 70569 Stuttgart, Germany}
\affiliation{Department of Physics, University of Tokyo, Bunkyo-ku, Hongo 7-3-1, Tokyo 113-0033, Japan}
\affiliation{Institute for Functional Matter and Quantum Technologies, University of Stuttgart, 70550 Stuttgart, Germany}
\author{M.~Hepting}
\email[]{hepting@fkf.mpg.de}
\affiliation{Max-Planck-Institute for Solid State Research, Heisenbergstra$\beta$e 1, 70569 Stuttgart, Germany}
\author{B.~Keimer}
\email[]{b.keimer@fkf.mpg.de}
\affiliation{Max-Planck-Institute for Solid State Research, Heisenbergstra$\beta$e 1, 70569 Stuttgart, Germany}

\date{\today}

\begin{abstract}
We investigate the magnetic ground state of single crystals of the ruthenium-dimer-based hexagonal perovskite \BZRO using magnetic susceptibility and resonant inelastic x-ray scattering (RIXS) measurements. 
While a previous study on powder samples exhibited intriguing magnetic behavior, questions about whether the spin state within a Ru$_{2}$O$_{9}$ dimer is a conventional $S = 3/2$ dimer or an orbital-selective $S = 1$ dimer were raised.
The RIXS spectra reveal magnetic excitations from Hund's intraionic multiplet and intradimer spin-triplet transitions.
The observed transition energies of the Hund's intraionic multiplets align with the $S=3/2$ ground state, contrasting with the theoretically proposed orbital-selective $S=1$ dimer state.
High-temperature magnetic susceptibility analysis confirms the realization of the spin $S=3/2$ dimer state, and the extracted intradimer coupling is consistent with the spin-triplet transition energy observed in the RIXS spectra.
These results highlights the ability of ``spectroscopic fingerprinting'' by RIXS to determine the magnetic ground states of complex materials.
\end{abstract}

\maketitle



\section{Introduction} 
\label{sec:introduction}
Hexagonal perovskites have served as an intriguing platform for quantum magnetism~\cite{Nguyen2021}.
Unlike cubic perovskites, where metal-oxygen octahedra share corners, hexagonal perovskites are characterized by their face-sharing octahedral frameworks, which can result in the formation of dimers, trimers, tetramers, or even longer chains.
The metal-metal distances in such networks are generally shorter (and the overlap of metal $d$-orbitals correspondingly larger) than those in networks of corner-sharing octahedra.
This can lead to unusual magnetic quantum correlations~\cite{Nguyen2021,Streltsov2016}.

Within this family, ruthenium-dimer-based hexagonal perovskites \BMRO ($M=$ cations), featuring Ru$_{2}$O$_{9}$ dimers, have been extensively studied.
The oxidation state of $M$ plays a crucial role in determining the valence state of the Ru atom, leading to a variety of magnetic states.
Magnetic cations such as Ni$^{2+}$, Co$^{2+}$, and rare-earth metal $R^{3+}$ ions induce long-range magnetic order~\cite{Lightfoot1990,Rijssenbeek1999,Doi2001,DoiJSSC2002,DoiJMC2002,Senn2013,Basu2018,Basu2019,Basu2020,Kushwaha2024}. 
Notably, in systems with rare-earth metal ions, \BRRO exhibits unusual magnetoelectric couplings~\cite{Basu2018,Basu2019,Kushwaha2024}.
On the other hand, nonmagnetic cations can also lead to diverse magnetic states in the Ru$_{2}$O$_{9}$ dimer.
For instance, when $M$ is a nonmagnetic divalent cation such as Mg$^{2+}$, Ca$^{2+}$, Cd$^{2+}$, or Sr$^{2+}$, the Ru$_{2}$O$_{9}$ dimer adopts a gapped nonmagnetic singlet state with a total spin of $S_{\rm tot}=0$~\cite{Darriet1976}.
In contrast, when $M$ is a nonmagnetic trivalent cation such as Y$^{3+}$, In$^{3+}$, Lu$^{3+}$, or La$^{3+}$, the Ru$_{2}$O$_{9}$ dimer exhibits either a molecular spin $S_{\rm tot} = 1/2$ state~\cite{Ziat2017} or an orbital-selective $S_{\rm tot}=3/2$ state~\cite{Chen2020}.

\begin{figure}[!t]
\includegraphics[scale=1]{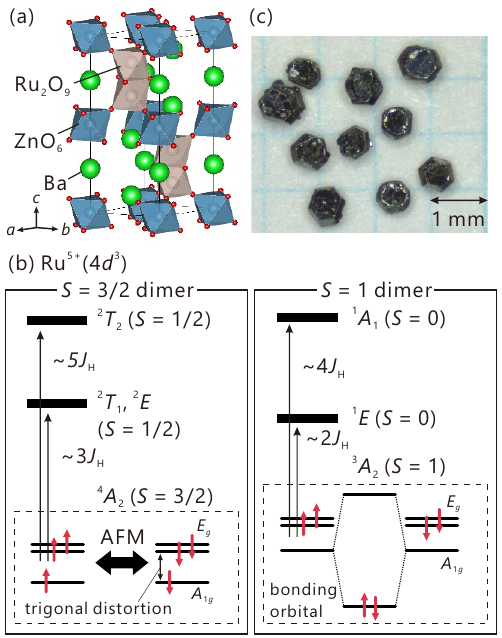}
\caption{(a) Crystal structure of \BZRO with hexagonal space group $P6_{3}/mmc$.
(b)  Energy levels of the $t_{2g}^{3}$ electron configurations for the Ru$^{5+}$ ion dimer, depicting the cases of the $S=3/2$ state (left) and the $S=1$ state (right), as discussed in the text. The Ru$^{5+}$ spins couple antiferromagnetically (AFM) within the dimer. $J_{\rm H}$ denotes the Hund coupling. 
(c) Typical \BZRO single crystals used in this work.  The typical mass of a crystal is approximately 1~mg.}
\label{fig:crystal}
\end{figure}

Among these, \BZRO, characterized by the nonmagnetic divalent Zn$^{2+}$ ion, stands out as one of the most intriguing dimer-based hexagonal perovskites~\cite{Terasaki2017,Yamamoto2018}.
Its crystal structure is almost hexagonal with the space group $P6_{3}/mmc$~\cite{Lightfoot1990}, but slight distortions lower the symmetry to monoclinic (space group $C2/c$)~\cite{Injac2020}. 
In this structure, the Ru$_{2}$O$_{9}$ dimers are separated by ZnO$_{6}$ octahedra and form a triangular lattice in the crystallographic $ab$ plane, as illustrated in Fig.~\ref{fig:crystal}(a).
The Zn$^{2+}$ ions lead to the Ru$^{5+}$ valence state with a $4d^{3}$ electronic configuration.
Electrical resistivity measurements confirm that \BZRO is highly insulating~\cite{Terasaki2017}.
Notably, \BZRO shows no evidence of long-range magnetic order or spin-singlet-gapped behavior down to 37 mK~\cite{Terasaki2017}, which contrasts with its analogues \BMRO ($M=$ Mg, Ca, Cd, and Sr) showing a singlet gapped behavior~\cite{Darriet1976}.
This observation indicates that the magnetic state of the Ru$_{2}$O$_{9}$ dimer in \BZRO may be unconventional.

However, the magnetic state of the isolated Ru$_{2}$O$_{9}$ dimer in \BZRO remains unresolved.
Figure~\ref{fig:crystal}(b) outlines two possible scenarios for the energy levels of the dimerized $t_{2g}^{3}$-$t_{2g}^{3}$ electron configuration of the Ru$^{5+}$ ions in a cubic crystal electric field with trigonal distortion.
The electronic structures can be characterized by the Hund's intraionic multiplets~\cite{Sugano1970}.
In \BZRO, the trigonal distortion splits the $t_{2g}$ orbitals into doubly degenerate $E_{g}$ orbitals and a nondegenerate $A_{1g}$ orbital.
In the conventional high-spin $t_{2g}^{3}$ state, each of the three spins occupies a separate orbital for the ground state, resulting in the high-spin $^{4}A_{2}$ multiplet with $S=3/2$, namely the $S=3/2$ spin dimer state [see the left panel of Fig.~\ref{fig:crystal}(b)].  
The associated low-spin excited states, $(^{2}T_{1}, {}^{2}E)$ and $^{2}T_{2}$, with $S=1/2$, appear at energies of $\sim 3J_{\rm H}$ and $\sim 5J_{\rm H}$ above the ground state, where $J_{\rm H}$ is the Hund coupling.
Alternatively, a theoretical study~\cite{Tanaka2020} proposes an orbital-selective $S=1$ spin dimer state in \BZRO.
In this scenario, strong hybridization of the $A_{1g}$ orbitals in the dimerized Ru$^{5+}$ ions can induce a singlet pair of spins per dimer in the lowest bonding orbital state, reducing the spin moment of each Ru$^{5+}$ ion to $S = 1$ at the $^{3}A_{1}$ multiplet, thus yielding the $S=1$ spin dimer~\cite{Tanaka2020,Streltsov2017} [see the right panel in Fig.~\ref{fig:crystal}(b)].
Accounting for the remaining doubly degenerate $E_{g}$ orbital and two spins, the associated low-spin excited multiplets include an $^{1}E$ state with $S=0$ at $\sim2J_{\rm H}$ and an $^{1}A_{1}$ state with $S=0$ at $\sim4J_{\rm H}$ above the ground state.

To determine the magnetic state of the Ru$_{2}$O$_{9}$ dimer in \BZRO, we undertake a comprehensive investigation of the magnetic properties of single-crystals using magnetic susceptibility and resonant inelastic x-ray scattering (RIXS) measurements.
The high-temperature magnetic susceptibility and the RIXS spectra provide evidence supporting the $S=3/2$ dimer rather than the $S=1$ dimer scenario.

\section{Experimental details}\label{sec:experimental_details}
Hexagonally shaped single crystals of \BZRO [Fig.~\ref{fig:crystal}(c)] were grown using the typical PbO flux method (see the Appendix for more details).
The crystal quality was characterized by x-ray diffraction (XRD) and electron-dispersive x-ray (EDX) measurements, which identified the presence of an impurity phase, the metallic hexagonal perovskite 4H-BaRuO$_{3}$~\cite{Hong1997} (see the detailed characterization in the Appendix).

RIXS measurements were performed at beamline P01 at the PETRA-III synchrotron at DESY, Germany, using the IRIXS spectrometer~\cite{Gretarsson2020}. 
All RIXS spectra were collected at $T=296$ and 15~K with an effective energy resolution of $\Delta E=75$~meV (full width at half-maximum) and a scattering angle of $90^{\circ}$. The RIXS studies were carried out in the $(H,0,L)$ scattering geometry.
The momentum coordinates are quoted in terms of reciprocal lattice units (r.l.u.), based on the lattice constants listed in Table~\ref{tb:xtal_xrd} of the Appendix.

Magnetic characterization was conducted using a vibrating-sample magnetometer (VSM) within a Quantum Design Magnetic Property Measurement System (MPMS).
High-temperature magnetic susceptibility measurements ($T \geq 400$ K) were conducted with a Quantum Design oven option.

\section{Results and discussion}
\subsection{RIXS}
\label{sec:RIXS}
\begin{figure}[tbp]
\includegraphics[scale=1]{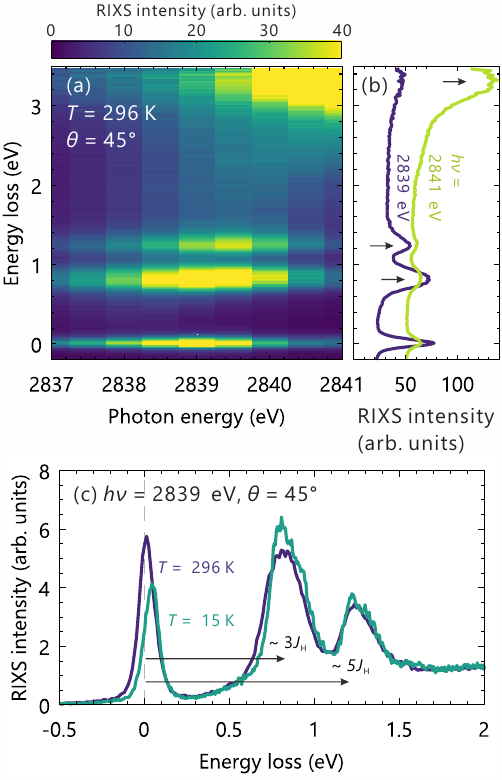}
\caption{(a) RIXS intensity map of the \BZRO crystal for incident photon energies varied across the Ru $L_{3}$-edge, taken at $\theta=45^{\circ}$ and at $T=296$~K. (b) RIXS spectra for incident photon energies of 2839 and 2841~eV. The data are horizontally offset for visibility.  (c) Ru $L_{3}$ RIXS spectra at $T=15$ and 296~K, measured with an incident angle $\theta=45^{\circ}$. The incident angle $\theta$ is defined as the angle between the incident photon $\mathbf{k}_{\rm i}$ and the sample surface. Arrows in (b) and (c) indicate the $t_{2g}\rightarrow e_{g}$ excitation and Hund's intraionic multiplet excitations discussed in the text.}
\label{fig:RIXS_highE}
\end{figure}

Figure~\ref{fig:RIXS_highE}(a) shows RIXS spectra of a \BZRO crystal, with the incident photon energy varied across the Ru $L_{3}$-edge.
Distinct resonance peaks are observed at approximately 1 and 3~eV, which are attributed to $dd$ excitations.
Two sharp features at 0.8 and 1.2~eV [Fig.~\ref{fig:RIXS_highE}(b)] correspond to the Hund's intraionic multiplet transitions within the $t_{2g}$ orbital manifold as discussed further below.
The enhanced feature near 3.3~eV [Fig.~\ref{fig:RIXS_highE}(b)] is attributed to $t_{2g}\rightarrow e_{g}$ excitations, similar to prior observations in other ruthenates, Ca$_{2}$RuO$_{4}$~\cite{Gretarsson2019} and Ca$_{3}$Ru$_{2}$O$_{7}$~\cite{Bertinshaw2021}.
Notably, these sharp orbital excitations indicate the presence of a highly insulating state, contrasting with electronic continuum excitations observed in metallic systems such as Sr$_{2}$RuO$_{4}$~\cite{Suzuki2023} and Ca$_{3}$Ru$_{2}$O$_{7}$~\cite{Bertinshaw2021}.
Thus, the measured RIXS spectra predominantly reflect the primary insulating \BZRO phase~\cite{Terasaki2017}, with no evidence for contributions from the metallic impurity phase of 4H-BaRuO$_{3}$~\cite{Zhao2007}. 
As the RIXS penetration depth is below 1~$\mu$m, this is consistent with the observation that the near-surface region of the crystals is composed of the \BZRO majority phase, whereas the impurity phase is concentrated in the crystal interior (see the Appendix).

Figure~\ref{fig:RIXS_highE}(c) shows the RIXS spectra with an incident photon energy at the Ru $L_{3}$ edge (2839~eV).
The two prominent peaks at 0.80 and 1.22~eV correspond to Hund's intraionic $t_{2g}^{3}$ multiplet transitions.
Based on the transition energy ratio $1.22/0.80 = 1.53$, these peaks are assigned to the $\sim3J_{\rm H}$ and $\sim5J_{\rm H}$ transitions from the $^{4}A_{2}$ multiplet, rather than the $\sim2J_{\rm H}$ and $\sim4J_{\rm H}$ transitions expected from the $^{3}A_{2}$ multiplet, as depicted in Fig.~\ref{fig:crystal}(b).
These high-energy excitations thus provide clear evidence for the $S=3/2$ spin state as the ground state in the Ru$_{2}$O$_{9}$ dimer of \BZRO.
The estimated Hund coupling, $J_{\rm H}\sim0.25$~eV, agrees with the value reported in another Ru$^{5+}$ compound, SrRu$_{2}$O$_{6}$~\cite{Suzuki2019}. 

\begin{figure}[!t]
\includegraphics[scale=1]{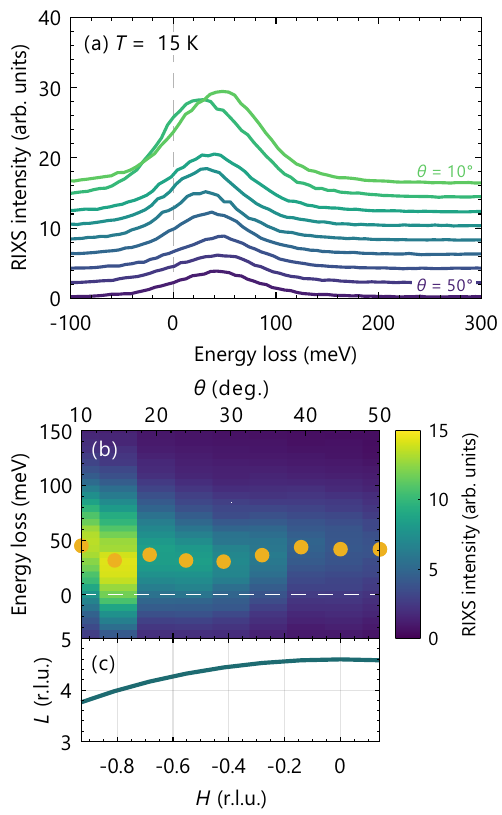}
\caption{(a) Incident angle $\theta$ dependent RIXS spectra at 15~K. The spectra were acquired from $\theta=10^{\circ}$ to $50^{\circ}$ with a $5^{\circ}$ step. (b) Color map of the RIXS intensities in (a). Yellow circles denote peak positions extracted by Gaussian fitting. (c) Measured positions of the RIXS spectra in reciprocal lattice units $(H, 0, L)$.}
\label{fig:RIXS_lowE}
\end{figure}

A noticeable peak is also observed in the quasielastic region below 0.05~eV [see Fig.~\ref{fig:RIXS_highE}(c)].
This peak displays clear resonance behavior as a function of the incident photon energy [Fig.~\ref{fig:RIXS_highE}(a)], and its center shifts to higher energies at low temperatures [Fig.~\ref{fig:RIXS_highE}(c)], indicating an inelastic origin.
The observed inelastic signal can be, in principle, attributed to either phonons or magnetic excitations.
However, phonons typically couple more strongly with the $e_{g}$ orbital resonance than with the $t_{2g}$ orbital resonance, as seen in cuprates~\cite{Braicovich2020}, because $e_{g}$ orbitals extend towards ligands, and their energies are expected to be approximately $T$-independent. 
Since we selected the $t_{2g}$ orbital resonance and observed a pronounced blue-shift upon cooling, phonon contributions are expected to be minimal.
Thus, the inelastic peak is primarily attributed to magnetic excitations, and the observed shift at 15~K suggests the development of spin-spin correlations in \BZRO.


To investigate potential dispersive magnetic excitations, we measured the incident-angle $\theta$ dependence of the RIXS spectra at 15~K, as shown in Fig.~\ref{fig:RIXS_lowE}(a).
Resolution-limited peaks are observed in the 30--45~meV energy range, with an average value of 37(13)~meV, which we can now assign to a conventional spin-triplet excitation in the antiferromagnetic $S=3/2$ dimer system.
In the spin-triplet excitation, the central energy of the band typically reflects the intradimer coupling~\cite{Zapf2014}, allowing us to estimate the strength of this coupling in \BZRO to be approximately 37(13)~meV.
As result, the observed Hund's intraionic multiplets and spin-triplet excitations indicate that the magnetic state in \BZRO is characterized by the $S=3/2$ dimer configuration.

A dispersive feature of the excitation is not clearly discernible in our experiment [see Fig.~\ref{fig:RIXS_lowE}(b)] even though the measured incident angles cover a wide range in the momentum plane $(H,0,L)$ [Fig.~\ref{fig:RIXS_lowE}(c)].
This suggests that the interdimer coupling is relatively weak, as expected in view of the large spatial separation of the dimers [Fig.~\ref{fig:crystal}(a)].
A more detailed analysis would require high-resolution spectroscopic measurements, such as inelastic neutron scattering.  


\subsection{Magnetic susceptibility}
\label{sec:susceptibility}

\begin{figure}[!t]
\includegraphics[scale=1]{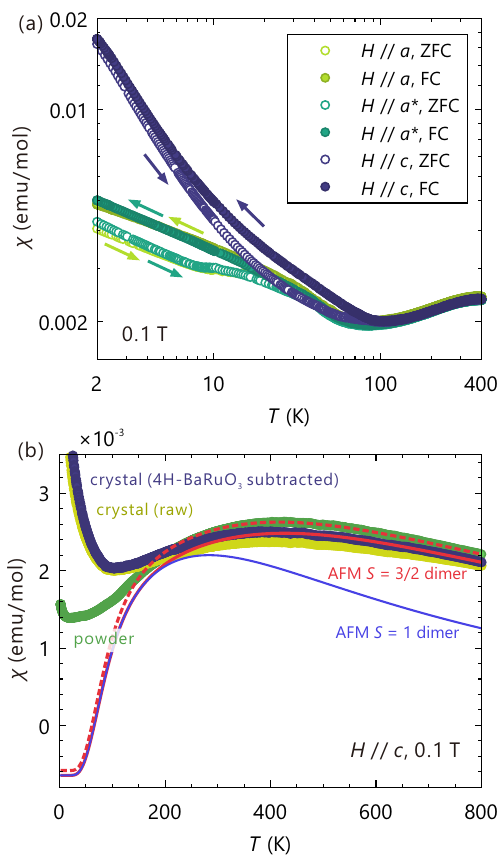}
\caption{(a) Magnetic susceptibility measured with a 0.1~T magnetic field applied along the $a$, $a^{*}$, and $c$ axes, with both axes on logarithmic scales. Open and filled circles denote zero-field cooling (ZFC) and field cooling (FC) data, respectively. Arrows indicate ZFC and FC processes. Note that a kink around 10~K in the ZFC data for the $\mathbf{H}\parallel\mathbf{a}$ and $\mathbf{a^{*}}$ might be an extrinsic effect related to a change in the temperature sweep rate at 10~K.
(b) High-temperature magnetic susceptibility measured with a 0.1~T magnetic field.  Yellow symbols represent the raw susceptibility of the \BZRO single crystal with the field applied along the $c$ axis, while the dark-blue symbols show the susceptibility after subtracting the constant Pauli paramagnetic contribution from the 4H-BaRuO$_{3}$ impurity phase. Green symbols correspond to the susceptibility of \BZRO powder. The red and blue curves represent the calculated susceptibilities of the antiferromagnetic (AFM) spin dimer models as described in the text.}
\label{fig:susceptibility}
\end{figure}

Figure~\ref{fig:susceptibility}(a) shows the temperature dependence of the magnetic susceptibility in the \BZRO single-crystals.
Below 100~K, pronounced anisotropic upturns with temperature hysteresis are observed, likely due to the freezing of free spins  within the sample and/or the isolated Ru spins occupying Zn sites~\cite{Yamamoto2018}.
This behavior indicates that the low-temperature susceptibility is heavily influenced by disorder such as a site-mixing between the Ru and Zn atoms, as observed in the single-crystal XRD (see the Appendix).
Thus, we focus on the high-temperature magnetic properties, which are unaffected by weakly coupled spins, to elucidate the intrinsic magnetic state in \BZRO.

Figure~\ref{fig:susceptibility}(b) shows the magnetic susceptibility of \BZRO, measured up to 800~K.
Above 400~K, measurements were conducted only for the $\mathbf{H}\parallel\mathbf{c}$ configuration, since the susceptibility becomes isotropic above 100~K [see Fig.~\ref{fig:susceptibility}(a)].
A broad maximum appears around 400~K, followed by a steady decrease at higher temperatures, indicating a crossover from developing antiferromagnetic spin correlations to a paramagnetic Curie tail in the high-temperature regime.

To distinguish between the primary \BZRO and impurity 4H-BaRuO$_{3}$ phases, we subtract the magnetic contribution of 4H-BaRuO$_{3}$, accounting for the $4:1$ volume fraction extracted from the XRD measurement on the pulverized crystals [see Fig.~\ref{fig:XRD_EDX}(a) in the Appendix].
The metallic 4H-BaRuO$_{3}$ phase exhibits exchange-enhanced Pauli paramagnetic behavior at low temperatures, and its susceptibility weakly decreases at high temperatures due to weakening exchange interactions between conduction electrons~\cite{Zhao2007}.
As a first order approximation, we assume that this susceptibility decline is minimal, and we extrapolate a constant Pauli paramagnetic value, $\chi_{\rm Pauli}=6.58\times 10^{-2}$~emu/mol(BaRuO$_{3}$)~\cite{Zhao2007}, to higher temperatures.
The corrected susceptibility data [dark-blue symbols in Fig.~\ref{fig:susceptibility}(b)] agrees well with single-phase powder data at high temperatures [green symbols in Fig.~\ref{fig:susceptibility}(b)], with a minor offset that may be attributable to diamagnetic effects.

The magnetic state in the Ru$_{2}$O$_{9}$ dimer is analyzed using the corrected susceptibility.
We calculate the susceptibility based on an isolated Heisenberg antiferromagnetic spin dimer model.
In an antiferromagnetic spin dimer with $S=3/2$, the magnetic states within a dimer are labeled by the total spins $S_{\rm tot}=0,1,2$, and $3$.
Their energies are given by $E = \frac{J}{2}\left[S_{\rm tot}(S_{\rm tot}+1)-2S(S+1)\right]$, where $J$ is the intradimer coupling. 
The uniform magnetic susceptibility per dimer is then represented by
\begin{eqnarray}
&\chi& = \frac{2N_{\rm A}}{\kbT}(g\mu_{\rm B})^{2} \nonumber \\
&&\times \frac{e^{-J/\kbT}+5e^{-3J/\kbT}+14e^{-6J/\kbT}}{1+3e^{-J/\kbT}+5e^{-3J/\kbT}+7e^{-6J/\kbT}},
\label{eq:MF_chi_3}
\end{eqnarray}
where $N_{\rm A}$ is Avogadro's number, $g$ is the $g$-factor, and $\mu_{\rm B}$ is Bohr magneton.   
For simplicity, we fix $g=2$ and refine the intradimer coupling $J$.
Additionally, we introduce a diamagnetic term $\chi_{0}$ for a constant background in the fitting.
Assuming that the contribution from free-spin paramagnetic impurities is weak and $T$-independent in the high-temperature limit, we fit the data above 400~K. 
The resulting curves are in good agreement with the high-temperature susceptibility data for the single-crystal sample [solid red curve in Fig.~\ref{fig:susceptibility}(b)] and the phase pure powder sample [dashed red curve in Fig.~\ref{fig:susceptibility}(b)]. 
The refined parameters are $J=277(37)$~K and $\chi_{0}=-0.65(28)\times 10^{-3}$~emu/mol for the crystal, and $J=270(27)$~K and $\chi_{0}=-0.58(6)\times 10^{-3}$~emu/mol for the powder.
The antiferromagnetic $J$ value of $277(37)$~K [$=24(3)$~meV] reasonably aligns with $37(13)$~meV obtained from the RIXS measurement in Sec.~\ref{sec:RIXS} within the error of the measurements.

Similarly, we compare the susceptibility data with the antiferromagnetic spin $S=1$ dimer, where the magnetic states are labeled by $S_{\rm tot} = 0,1$, and $2$. 
The calculated susceptibility is given by
\begin{equation}
\chi = \frac{2N_{\rm A}}{\kbT}(g\mu_{\rm B})^{2}
\frac{e^{-J/\kbT}+5e^{-3J/\kbT}}{1+3e^{-J/\kbT}+5e^{-3J/\kbT}}.
\label{eq:MF_chi_1}
\end{equation}
The reference curve for the $S=1$ dimer susceptibility with $J=277$~K and $\chi_{0}=-0.65\times 10^{-3}$~emu/mol [blue curve in Fig.~\ref{fig:susceptibility}(b)] shows a more pronounced decrease at high temperatures, inconsistent with the observed susceptibility trend.
Consequently, the high-temperature magnetic susceptibility data support the conclusion that the magnetic state in \BZRO is well described by the $S=3/2$ dimer rather than the $S=1$ dimer.

\section{Conclusions}\label{sec:conclusions}
Our magnetic and RIXS measurements provide additional insights into the magnetic ground state of the ruthenium-dimer-based hexagonal perovskite \BZRO. 
RIXS spectra reveal the Hund's intraionic multiplets and the spin-triplet excitation associated with the $S=3/2$ dimer, providing strong evidence for the realization of the $S=3/2$ spin dimer  in \BZRO rather than the theoretically proposed orbital-selective $S=1$ dimer.
Our high-temperature magnetic susceptibility analysis confirms the presence of the antiferromagnetic $S=3/2$ dimer.
For future perspective, improving the quality of the single crystals will be crucial for exploring the low-energy magnetic properties in more detail, as the current sample is affected by site-mixing between Zn and Ru. 

More generally, our study illustrates the power of RIXS to elucidate the ground state of individual magnetic ions in complex solids by "spectroscopic fingerprinting".
At the same time, the resolution of modern RIXS spectrometers (including the IRIXS spectrometer for $4d$-electron compounds used here) suffices to determine the sign and strength of exchange interactions between magnetic ions in a wide variety of arrangements, including simple dimers in the current case and magnetic lattices giving rise to collective magnon excitations in other compounds.

\vspace{\baselineskip}
Experimental data associated with this manuscript are available from Ref.~\cite{dataset}.

\begin{acknowledgements}
We thank A.~Schnyder for helpful discussions, and E.~Br{\"u}cher and R.~K.~Kremer for technical support of heat capacity measurements.
The RIXS experiments were carried out at the beamline P01 of PETRA III at DESY.
We acknowledge funding from the Deutsche Forschungsgemeinschaft (DFG, German Research Foundation) – TRR 360 – 492547816 and from the European Research Council (ERC) under Advanced Grant No. 101141844 (SpecTera).
\end{acknowledgements}

\appendix*
\section{Crystal characterization}\label{sec:characterization}
\begin{table}[!b]
\caption{Structural parameters at 296~K from the single-crystal x-ray diffraction measurement. The data were refined using space group $P6_{3}/mmc$ with $a = b = 5.7682(7)$~{\AA}, $c=14.188(3)$~{\AA}, $\alpha = \beta = 90^{\circ}$, and $\gamma=120^{\circ}$. An agreement factor $S$ of the refinement is $S=1.41$.}
\begin{tabular}{lllll}
\hline \hline
 & $x$ & $y$ & $z$ & Occupancy \\ \hline
Ba1 \quad & 0 \quad & 0 \quad & 1/4\quad & 1\quad \\
Ba1 & 1/3 & 2/3 & 0.91045(3) & 1 \\
Zn1 & 0 & 0 & 0 & 0.91(2) \\
Ru1 & 0 & 0 & 0 & 0.09(2) \\
Ru2 & 1/3 & 2/3 & 0.15577(4) & 1 \\
O1 & 0.4849(5) & 0.9698(9) & 1/4 & 1 \\
O2 \qquad & 0.1702(4) \qquad & 0.3404(8) \qquad & 0.4164(2) \qquad & 1 \\ \hline
\end{tabular}
\label{tb:xtal_xrd}
\end{table}

Single crystal samples of \BZRO were synthesized by the following PbO flux method.
Initially \BZRO powder was synthesized from starting materials, BaCO$_{3}$, ZnO, and RuO$_{2}$.
Stoichiometric amounts of these materials were mixed and heated at 1200~$^{\circ}$C for over 54~hours.
A mixture of 0.5~g of \BZRO powder and PbO flux in a mass ratio of $1:15$ was placed in a platinum crucible, which was then enclosed in an alumina crucible.
The crucible was heated at 1100~$^{\circ}$C for 10 hours and slowly cooled to 900~$^{\circ}$C for 96 hours.
To grow larger crystals, this process was repeated at least four times, with additional PbO flux to compensate for its evaporation. 
The flux was subsequently removed using KOH solutions.

\begin{figure}[!t]
\includegraphics[scale=1]{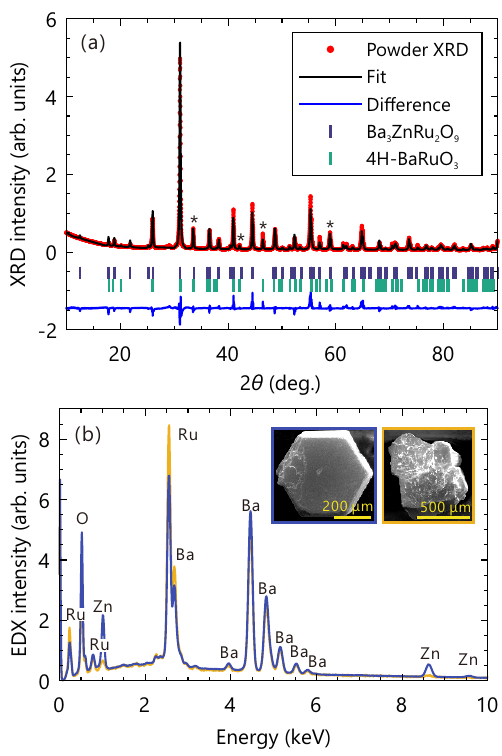}
\caption{(a) Powder XRD pattern of pulverized \BZRO crystals. A solid black line is the fit obtained from a Rietveld refinement, which includes the \BZRO phase (upper blue symbols) and a minority 4H-BaRuO$_{3}$ phase (lower
green symbols). Asterisks mark the most notable Bragg peaks of the latter phase.
(b) EDX spectra obtained from the surface (blue) and the interior (yellow) of an as-grown \BZRO crystal. The insets show SEM images of the crystals.}
\label{fig:XRD_EDX}
\end{figure}

Single crystal x-ray diffraction (XRD) measurements were conducted at 296~K using a SMART APEX-I CCD x-ray diffractometer (Bruker AXS) with monochromated Mo-K$\alpha$ radiation ($\lambda=0.71073$~{\AA}). 
The structure was refined using full matrix least-squares fitting with the SHELXTL software package~\cite{Sheldrick2008,Sheldrick2014}.
Structural parameters of the \BZRO crystals were determined by single-crystal XRD.
Note that we refined all the reflections using the space group $P6_{3}/mmc$ because the monoclinic distortions are very subtle~\cite{Injac2020}.
We tested for site-mixing at the Zn site with the Ru atom. 
The refinement results are presented in Table.~\ref{tb:xtal_xrd}.
Although the refined structural parameters are consistent with the previous report~\cite{Lightfoot1990}, 9{\%} site mixing at the Zn site is observed.
Including the site-mixing parameter slightly improves the agreement factor $S$ of the refinement from $S=1.43$ without the site-mixing to $S=1.41$ with the site-mixing.

\begin{figure}[!t]
\includegraphics[scale=1]{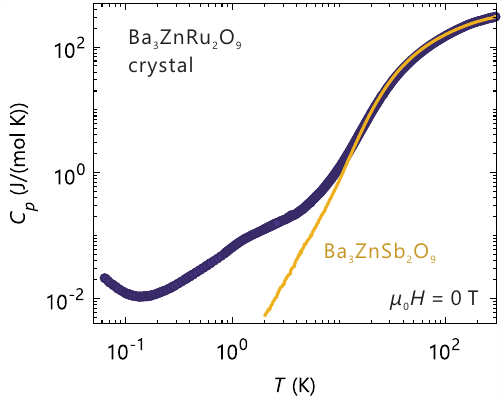}
\caption{Specific heat of \BZRO in zero magnetic field, plotted on logarithmic scales for both axes. The solid curve represents the phonon contribution, measured using the nonmagnetic reference Ba$_{3}$ZnSb$_{2}$O$_{9}$ powder sample.}
\label{fig:specific_heat}
\end{figure}

Powder XRD measurements on pulverized \BZRO were conducted using a Rigaku Miniflex diffractometer with Cu $K\alpha$ radiation. 
The data were analyzed using the Rietveld method using FullProf software~\cite{FullProf}.
The powder XRD results reveal the presence of additional disorder within the single crystal.
A hexagonal perovskite phase, 4H-BaRuO$_{3}$~\cite{Hong1997}, is identified [Fig.~\ref{fig:XRD_EDX}(a)].
4H-BaRuO$_{3}$ crystallizes in the space group $P6_{3}/mmc$ and consists of Ru$_{2}$O$_{9}$ dimers similar to those in \BZRO.
The volume fraction of \BZRO and 4H-BaRuO$_{3}$ is estimated to be approximately $4:1$ based on the diffraction pattern. 
Note that the site mixing observed in the single-crystal XRD is not resolvable from the powder XRD data.

Energy-dispersive x-ray (EDX) measurements were performed using a NORAN System 7 (NSS212E) detector in a Tescan Vega (TS-5130MM) scanning electron microscope (SEM) to compare the chemical compositions between the surface and interior of the \BZRO crystals.
Figure~\ref{fig:XRD_EDX}(b) shows EDX spectra from both the crystal surface and interior.
The spectrum taken from the crystal surface shows a stoichiometry of Ba$_{3.1(1)}$Zn$_{0.9(1)}$Ru$_{2.2(1)}$O$_{8.9(2)}$, which aligns perfectly with \BZRO.
In contrast, the spectrum from the crystal interior reveals a suppression of the Zn peaks and enhancement of the Ru peaks, resulting in a stoichiometry of Ba$_{3.2(1)}$Zn$_{0.1(1)}$Ru$_{3.1(1)}$O$_{8.6(2)}$.
This suggests that the crystal surface is predominantly composed of the primary phase \BZRO, while the impurity phase 4H-BaRuO$_{3}$ is more concentrated within the crystal interior.

Heat capacity was measured by standard relaxation calorimetry.
The measurements were conducted using a Quantum Design Physical Property Measurement System (PPMS) for temperatures above 2~K and using a $^{3}$He-$^{4}$He dilution refrigerator for temperatures below 2~K. 
The specific heat data across a wide temperature range at zero magnetic field are shown in Fig.~\ref{fig:specific_heat}.
Above 10~K, the data closely aligns with those of the nonmagnetic reference material Ba$_{3}$ZnSb$_{2}$O$_{9}$, while the magnetic contribution becomes apparent below 10~K, consistent with the previous powder study~\cite{Terasaki2017}.



\begin{thebibliography}{34}%
\makeatletter
\providecommand \@ifxundefined [1]{%
 \@ifx{#1\undefined}
}%
\providecommand \@ifnum [1]{%
 \ifnum #1\expandafter \@firstoftwo
 \else \expandafter \@secondoftwo
 \fi
}%
\providecommand \@ifx [1]{%
 \ifx #1\expandafter \@firstoftwo
 \else \expandafter \@secondoftwo
 \fi
}%
\providecommand \natexlab [1]{#1}%
\providecommand \enquote  [1]{``#1''}%
\providecommand \bibnamefont  [1]{#1}%
\providecommand \bibfnamefont [1]{#1}%
\providecommand \citenamefont [1]{#1}%
\providecommand \href@noop [0]{\@secondoftwo}%
\providecommand \href [0]{\begingroup \@sanitize@url \@href}%
\providecommand \@href[1]{\@@startlink{#1}\@@href}%
\providecommand \@@href[1]{\endgroup#1\@@endlink}%
\providecommand \@sanitize@url [0]{\catcode `\\12\catcode `\$12\catcode
  `\&12\catcode `\#12\catcode `\^12\catcode `\_12\catcode `\%12\relax}%
\providecommand \@@startlink[1]{}%
\providecommand \@@endlink[0]{}%
\providecommand \url  [0]{\begingroup\@sanitize@url \@url }%
\providecommand \@url [1]{\endgroup\@href {#1}{\urlprefix }}%
\providecommand \urlprefix  [0]{URL }%
\providecommand \Eprint [0]{\href }%
\providecommand \doibase [0]{https://doi.org/}%
\providecommand \selectlanguage [0]{\@gobble}%
\providecommand \bibinfo  [0]{\@secondoftwo}%
\providecommand \bibfield  [0]{\@secondoftwo}%
\providecommand \translation [1]{[#1]}%
\providecommand \BibitemOpen [0]{}%
\providecommand \bibitemStop [0]{}%
\providecommand \bibitemNoStop [0]{.\EOS\space}%
\providecommand \EOS [0]{\spacefactor3000\relax}%
\providecommand \BibitemShut  [1]{\csname bibitem#1\endcsname}%
\let\auto@bib@innerbib\@empty
\bibitem [{\citenamefont {Nguyen}\ and\ \citenamefont
  {Cava}(2021)}]{Nguyen2021}%
  \BibitemOpen
  \bibfield  {author} {\bibinfo {author} {\bibfnamefont {L.~T.}\ \bibnamefont
  {Nguyen}}\ and\ \bibinfo {author} {\bibfnamefont {R.~J.}\ \bibnamefont
  {Cava}},\ }\bibfield  {title} {\bibinfo {title} {{Hexagonal Perovskites as
  Quantum Materials}},\ }\href {https://doi.org/10.1021/acs.chemrev.0c00622}
  {\bibfield  {journal} {\bibinfo  {journal} {Chem. Rev.}\ }\textbf {\bibinfo
  {volume} {121}},\ \bibinfo {pages} {2935} (\bibinfo {year}
  {2021})}\BibitemShut {NoStop}%
\bibitem [{\citenamefont {Streltsov}\ and\ \citenamefont
  {Khomskii}(2016)}]{Streltsov2016}%
  \BibitemOpen
  \bibfield  {author} {\bibinfo {author} {\bibfnamefont {S.~V.}\ \bibnamefont
  {Streltsov}}\ and\ \bibinfo {author} {\bibfnamefont {D.~I.}\ \bibnamefont
  {Khomskii}},\ }\bibfield  {title} {\bibinfo {title} {{Covalent bonds against
  magnetism in transition metal compounds}},\ }\href
  {https://doi.org/10.1073/pnas.1606367113} {\bibfield  {journal} {\bibinfo
  {journal} {Proc. Natl. Acad. Sci. U.S.A.}\ }\textbf {\bibinfo {volume} {113}},\
  \bibinfo {pages} {10491} (\bibinfo {year} {2016})}\BibitemShut {NoStop}%
\bibitem [{\citenamefont {Lightfoot}\ and\ \citenamefont
  {Battle}(1990)}]{Lightfoot1990}%
  \BibitemOpen
  \bibfield  {author} {\bibinfo {author} {\bibfnamefont {P.}~\bibnamefont
  {Lightfoot}}\ and\ \bibinfo {author} {\bibfnamefont {P.}~\bibnamefont
  {Battle}},\ }\bibfield  {title} {\bibinfo {title} {{The crystal and magnetic
  structures of Ba$_{3}$NiRu$_{2}$O$_{9}$, Ba$_{3}$CoRu$_{2}$O$_{9}$, and
  Ba$_{3}$ZnRu$_{2}$O$_{9}$}},\ }\href
  {https://doi.org/https://doi.org/10.1016/0022-4596(90)90309-L} {\bibfield
  {journal} {\bibinfo  {journal} {J. Solid State Chem.}\ }\textbf {\bibinfo
  {volume} {89}},\ \bibinfo {pages} {174} (\bibinfo {year} {1990})}\BibitemShut
  {NoStop}%
\bibitem [{\citenamefont {Rijssenbeek}\ \emph {et~al.}(1999)\citenamefont
  {Rijssenbeek}, \citenamefont {Jin}, \citenamefont {Zadorozhny}, \citenamefont
  {Liu}, \citenamefont {Batlogg},\ and\ \citenamefont
  {Cava}}]{Rijssenbeek1999}%
  \BibitemOpen
  \bibfield  {author} {\bibinfo {author} {\bibfnamefont {J.~T.}\ \bibnamefont
  {Rijssenbeek}}, \bibinfo {author} {\bibfnamefont {R.}~\bibnamefont {Jin}},
  \bibinfo {author} {\bibfnamefont {Y.}~\bibnamefont {Zadorozhny}}, \bibinfo
  {author} {\bibfnamefont {Y.}~\bibnamefont {Liu}}, \bibinfo {author}
  {\bibfnamefont {B.}~\bibnamefont {Batlogg}},\ and\ \bibinfo {author}
  {\bibfnamefont {R.~J.}\ \bibnamefont {Cava}},\ }\bibfield  {title} {\bibinfo
  {title} {{Electrical and magnetic properties of the two crystallographic
  forms of ${\mathrm{BaRuO}}_{3}$}},\ }\href
  {https://doi.org/10.1103/PhysRevB.59.4561} {\bibfield  {journal} {\bibinfo
  {journal} {Phys. Rev. B}\ }\textbf {\bibinfo {volume} {59}},\ \bibinfo
  {pages} {4561} (\bibinfo {year} {1999})}\BibitemShut {NoStop}%
\bibitem [{\citenamefont {Doi}\ \emph {et~al.}(2001)\citenamefont {Doi},
  \citenamefont {Hinatsu}, \citenamefont {Shimojo},\ and\ \citenamefont
  {Ishii}}]{Doi2001}%
  \BibitemOpen
  \bibfield  {author} {\bibinfo {author} {\bibfnamefont {Y.}~\bibnamefont
  {Doi}}, \bibinfo {author} {\bibfnamefont {Y.}~\bibnamefont {Hinatsu}},
  \bibinfo {author} {\bibfnamefont {Y.}~\bibnamefont {Shimojo}},\ and\ \bibinfo
  {author} {\bibfnamefont {Y.}~\bibnamefont {Ishii}},\ }\bibfield  {title}
  {\bibinfo {title} {{Crystal Structure and Magnetic Properties of
  6H-Perovskite Ba$_{3}$NdRu$_{2}$O$_{9}$}},\ }\href
  {https://www.sciencedirect.com/science/article/pii/S0022459601992965}
  {\bibfield  {journal} {\bibinfo  {journal} {J. Solid State Chem.}\ }\textbf
  {\bibinfo {volume} {161}},\ \bibinfo {pages} {113} (\bibinfo {year}
  {2001})}\BibitemShut {NoStop}%
\bibitem [{\citenamefont {Doi}\ \emph {et~al.}(2002)\citenamefont {Doi},
  \citenamefont {Matsuhira},\ and\ \citenamefont {Hinatsu}}]{DoiJSSC2002}%
  \BibitemOpen
  \bibfield  {author} {\bibinfo {author} {\bibfnamefont {Y.}~\bibnamefont
  {Doi}}, \bibinfo {author} {\bibfnamefont {K.}~\bibnamefont {Matsuhira}},\
  and\ \bibinfo {author} {\bibfnamefont {Y.}~\bibnamefont {Hinatsu}},\
  }\bibfield  {title} {\bibinfo {title} {{Crystal Structures and Magnetic
  Properties of 6H-Perovskites Ba$_{3}$MRu$_{2}$O$_{9}$ (M=Y, In, La, Sm, Eu,
  and Lu)}},\ }\href {https://doi.org/https://doi.org/10.1006/jssc.2002.9538}
  {\bibfield  {journal} {\bibinfo  {journal} {J. Solid State Chem.}\ }\textbf
  {\bibinfo {volume} {165}},\ \bibinfo {pages} {317} (\bibinfo {year}
  {2002})}\BibitemShut {NoStop}%
\bibitem [{\citenamefont {Doi}\ and\ \citenamefont
  {Hinatsu}(2002)}]{DoiJMC2002}%
  \BibitemOpen
  \bibfield  {author} {\bibinfo {author} {\bibfnamefont {Y.}~\bibnamefont
  {Doi}}\ and\ \bibinfo {author} {\bibfnamefont {Y.}~\bibnamefont {Hinatsu}},\
  }\bibfield  {title} {\bibinfo {title} {{Magnetic and calorimetric studies on
  Ba$_{3}$LnRu$_{2}$O$_{9}$ (Ln = Gd{,} Ho–Yb) with 6H-perovskite
  structure}},\ }\href {https://doi.org/10.1039/B111504A} {\bibfield  {journal}
  {\bibinfo  {journal} {J. Mater. Chem.}\ }\textbf {\bibinfo {volume} {12}},\
  \bibinfo {pages} {1792} (\bibinfo {year} {2002})}\BibitemShut {NoStop}%
\bibitem [{\citenamefont {Senn}\ \emph {et~al.}(2013)\citenamefont {Senn},
  \citenamefont {Kimber}, \citenamefont {Arevalo~Lopez}, \citenamefont {Hill},\
  and\ \citenamefont {Attfield}}]{Senn2013}%
  \BibitemOpen
  \bibfield  {author} {\bibinfo {author} {\bibfnamefont {M.~S.}\ \bibnamefont
  {Senn}}, \bibinfo {author} {\bibfnamefont {S.~A.~J.}\ \bibnamefont {Kimber}},
  \bibinfo {author} {\bibfnamefont {A.~M.}\ \bibnamefont {Arevalo~Lopez}},
  \bibinfo {author} {\bibfnamefont {A.~H.}\ \bibnamefont {Hill}},\ and\
  \bibinfo {author} {\bibfnamefont {J.~P.}\ \bibnamefont {Attfield}},\
  }\bibfield  {title} {\bibinfo {title} {{Spin orders and lattice distortions
  of geometrically frustrated 6H-perovskites
  Ba$_{3}{B}^{\ensuremath{'}}$Ru$_{2}$O$_{9}$ (${B}^{\ensuremath{'}}$ =
  La$^{3+}$, Nd$^{3+}$, and Y${}^{3+}$)}},\ }\href
  {https://doi.org/10.1103/PhysRevB.87.134402} {\bibfield  {journal} {\bibinfo
  {journal} {Phys. Rev. B}\ }\textbf {\bibinfo {volume} {87}},\ \bibinfo
  {pages} {134402} (\bibinfo {year} {2013})}\BibitemShut {NoStop}%
\bibitem [{\citenamefont {Basu}\ \emph {et~al.}(2018)\citenamefont {Basu},
  \citenamefont {Pautrat}, \citenamefont {Hardy}, \citenamefont {Loidl},\ and\
  \citenamefont {Krohns}}]{Basu2018}%
  \BibitemOpen
  \bibfield  {author} {\bibinfo {author} {\bibfnamefont {T.}~\bibnamefont
  {Basu}}, \bibinfo {author} {\bibfnamefont {A.}~\bibnamefont {Pautrat}},
  \bibinfo {author} {\bibfnamefont {V.}~\bibnamefont {Hardy}}, \bibinfo
  {author} {\bibfnamefont {A.}~\bibnamefont {Loidl}},\ and\ \bibinfo {author}
  {\bibfnamefont {S.}~\bibnamefont {Krohns}},\ }\bibfield  {title} {\bibinfo
  {title} {{Magnetodielectric coupling in a Ru-based 6H-perovskite,
  Ba$_{3}$NdRu$_{2}$O$_{9}$}},\ }\href {https://doi.org/10.1063/1.5034449}
  {\bibfield  {journal} {\bibinfo  {journal} {Appl. Phys. Lett.}\ }\textbf
  {\bibinfo {volume} {113}},\ \bibinfo {pages} {042902} (\bibinfo {year}
  {2018})}\BibitemShut {NoStop}%
\bibitem [{\citenamefont {Basu}\ \emph {et~al.}(2019)\citenamefont {Basu},
  \citenamefont {Caignaert}, \citenamefont {Ghara}, \citenamefont {Ke},
  \citenamefont {Pautrat}, \citenamefont {Krohns}, \citenamefont {Loidl},\ and\
  \citenamefont {Raveau}}]{Basu2019}%
  \BibitemOpen
  \bibfield  {author} {\bibinfo {author} {\bibfnamefont {T.}~\bibnamefont
  {Basu}}, \bibinfo {author} {\bibfnamefont {V.}~\bibnamefont {Caignaert}},
  \bibinfo {author} {\bibfnamefont {S.}~\bibnamefont {Ghara}}, \bibinfo
  {author} {\bibfnamefont {X.}~\bibnamefont {Ke}}, \bibinfo {author}
  {\bibfnamefont {A.}~\bibnamefont {Pautrat}}, \bibinfo {author} {\bibfnamefont
  {S.}~\bibnamefont {Krohns}}, \bibinfo {author} {\bibfnamefont
  {A.}~\bibnamefont {Loidl}},\ and\ \bibinfo {author} {\bibfnamefont
  {B.}~\bibnamefont {Raveau}},\ }\bibfield  {title} {\bibinfo {title}
  {{Enhancement of magnetodielectric coupling in 6H-perovskites
  ${\mathrm{Ba}}_{3}R{\mathrm{Ru}}_{2}{\mathrm{O}}_{9}$ for heavier rare-earth
  cations ($R=\mathrm{Ho},\mathrm{Tb}$)}},\ }\href
  {https://doi.org/10.1103/PhysRevMaterials.3.114401} {\bibfield  {journal}
  {\bibinfo  {journal} {Phys. Rev. Mater.}\ }\textbf {\bibinfo {volume} {3}},\
  \bibinfo {pages} {114401} (\bibinfo {year} {2019})}\BibitemShut {NoStop}%
\bibitem [{\citenamefont {Basu}\ \emph {et~al.}(2020)\citenamefont {Basu},
  \citenamefont {Caignaert}, \citenamefont {Damay}, \citenamefont {Heitmann},
  \citenamefont {Raveau},\ and\ \citenamefont {Ke}}]{Basu2020}%
  \BibitemOpen
  \bibfield  {author} {\bibinfo {author} {\bibfnamefont {T.}~\bibnamefont
  {Basu}}, \bibinfo {author} {\bibfnamefont {V.}~\bibnamefont {Caignaert}},
  \bibinfo {author} {\bibfnamefont {F.}~\bibnamefont {Damay}}, \bibinfo
  {author} {\bibfnamefont {T.~W.}\ \bibnamefont {Heitmann}}, \bibinfo {author}
  {\bibfnamefont {B.}~\bibnamefont {Raveau}},\ and\ \bibinfo {author}
  {\bibfnamefont {X.}~\bibnamefont {Ke}},\ }\bibfield  {title} {\bibinfo
  {title} {{Cooperative $\mathrm{Ru}(4d)\ensuremath{-}\mathrm{Ho}(4f)$ magnetic
  ordering and phase coexistence in the $6H$ perovskite multiferroic
  ${\mathrm{Ba}}_{3}\mathrm{Ho}{\mathrm{Ru}}_{2}{\mathrm{O}}_{9}$}},\ }\href
  {https://doi.org/10.1103/PhysRevB.102.020409} {\bibfield  {journal} {\bibinfo
   {journal} {Phys. Rev. B}\ }\textbf {\bibinfo {volume} {102}},\ \bibinfo
  {pages} {020409(R)} (\bibinfo {year} {2020})}\BibitemShut {NoStop}%
\bibitem [{\citenamefont {Kushwaha}\ \emph {et~al.}(2024)\citenamefont
  {Kushwaha}, \citenamefont {Roy}, \citenamefont {Kumar}, \citenamefont {dos
  Santos}, \citenamefont {Ghosh}, \citenamefont {Adroja}, \citenamefont
  {Caignaert}, \citenamefont {Perez}, \citenamefont {Pautrat},\ and\
  \citenamefont {Basu}}]{Kushwaha2024}%
  \BibitemOpen
  \bibfield  {author} {\bibinfo {author} {\bibfnamefont {E.}~\bibnamefont
  {Kushwaha}}, \bibinfo {author} {\bibfnamefont {G.}~\bibnamefont {Roy}},
  \bibinfo {author} {\bibfnamefont {M.}~\bibnamefont {Kumar}}, \bibinfo
  {author} {\bibfnamefont {A.~M.}\ \bibnamefont {dos Santos}}, \bibinfo
  {author} {\bibfnamefont {S.}~\bibnamefont {Ghosh}}, \bibinfo {author}
  {\bibfnamefont {D.~T.}\ \bibnamefont {Adroja}}, \bibinfo {author}
  {\bibfnamefont {V.}~\bibnamefont {Caignaert}}, \bibinfo {author}
  {\bibfnamefont {O.}~\bibnamefont {Perez}}, \bibinfo {author} {\bibfnamefont
  {A.}~\bibnamefont {Pautrat}},\ and\ \bibinfo {author} {\bibfnamefont
  {T.}~\bibnamefont {Basu}},\ }\bibfield  {title} {\bibinfo {title} {{Origin of
  spin-driven ferroelectricity and effect of external pressure on the complex
  magnetism of the $6H$ perovskite
  ${\mathrm{Ba}}_{3}\mathrm{Ho}{\mathrm{Ru}}_{2}{\mathrm{O}}_{9}$}},\ }\href
  {https://doi.org/10.1103/PhysRevB.109.224418} {\bibfield  {journal} {\bibinfo
   {journal} {Phys. Rev. B}\ }\textbf {\bibinfo {volume} {109}},\ \bibinfo
  {pages} {224418} (\bibinfo {year} {2024})}\BibitemShut {NoStop}%
\bibitem [{\citenamefont {Darriet}\ \emph {et~al.}(1976)\citenamefont
  {Darriet}, \citenamefont {Drillon}, \citenamefont {Villeneuve},\ and\
  \citenamefont {Hagenmuller}}]{Darriet1976}%
  \BibitemOpen
  \bibfield  {author} {\bibinfo {author} {\bibfnamefont {J.}~\bibnamefont
  {Darriet}}, \bibinfo {author} {\bibfnamefont {M.}~\bibnamefont {Drillon}},
  \bibinfo {author} {\bibfnamefont {G.}~\bibnamefont {Villeneuve}},\ and\
  \bibinfo {author} {\bibfnamefont {P.}~\bibnamefont {Hagenmuller}},\
  }\bibfield  {title} {\bibinfo {title} {{Interactions magn{\'e}tiques dans des
  groupements binucl{\'e}aires du Ruth{\'e}nium +V}},\ }\href
  {https://www.sciencedirect.com/science/article/pii/0022459676901705}
  {\bibfield  {journal} {\bibinfo  {journal} {J. Solid State Chem.}\ }\textbf
  {\bibinfo {volume} {19}},\ \bibinfo {pages} {213} (\bibinfo {year}
  {1976})}\BibitemShut {NoStop}%
\bibitem [{\citenamefont {Ziat}\ \emph {et~al.}(2017)\citenamefont {Ziat},
  \citenamefont {Aczel}, \citenamefont {Sinclair}, \citenamefont {Chen},
  \citenamefont {Zhou}, \citenamefont {Williams}, \citenamefont {Stone},
  \citenamefont {Verrier},\ and\ \citenamefont {Quilliam}}]{Ziat2017}%
  \BibitemOpen
  \bibfield  {author} {\bibinfo {author} {\bibfnamefont {D.}~\bibnamefont
  {Ziat}}, \bibinfo {author} {\bibfnamefont {A.~A.}\ \bibnamefont {Aczel}},
  \bibinfo {author} {\bibfnamefont {R.}~\bibnamefont {Sinclair}}, \bibinfo
  {author} {\bibfnamefont {Q.}~\bibnamefont {Chen}}, \bibinfo {author}
  {\bibfnamefont {H.~D.}\ \bibnamefont {Zhou}}, \bibinfo {author}
  {\bibfnamefont {T.~J.}\ \bibnamefont {Williams}}, \bibinfo {author}
  {\bibfnamefont {M.~B.}\ \bibnamefont {Stone}}, \bibinfo {author}
  {\bibfnamefont {A.}~\bibnamefont {Verrier}},\ and\ \bibinfo {author}
  {\bibfnamefont {J.~A.}\ \bibnamefont {Quilliam}},\ }\bibfield  {title}
  {\bibinfo {title} {{Frustrated spin-$\frac{1}{2}$ molecular magnetism in the
  mixed-valence antiferromagnets Ba$_{3}M$Ru$_{2}$O$_{9}$ ($M = $ In, Y,
  Lu)}},\ }\href {https://doi.org/10.1103/PhysRevB.95.184424} {\bibfield
  {journal} {\bibinfo  {journal} {Phys. Rev. B}\ }\textbf {\bibinfo {volume}
  {95}},\ \bibinfo {pages} {184424} (\bibinfo {year} {2017})}\BibitemShut
  {NoStop}%
\bibitem [{\citenamefont {Chen}\ \emph {et~al.}(2020)\citenamefont {Chen},
  \citenamefont {Verrier}, \citenamefont {Ziat}, \citenamefont {Clune},
  \citenamefont {Rouane}, \citenamefont {Bazier-Matte}, \citenamefont {Wang},
  \citenamefont {Calder}, \citenamefont {Taddei}, \citenamefont {dela Cruz},
  \citenamefont {Kolesnikov}, \citenamefont {Ma}, \citenamefont {Cheng},
  \citenamefont {Liu}, \citenamefont {Quilliam}, \citenamefont {Musfeldt},
  \citenamefont {Zhou},\ and\ \citenamefont {Aczel}}]{Chen2020}%
  \BibitemOpen
  \bibfield  {author} {\bibinfo {author} {\bibfnamefont {Q.}~\bibnamefont
  {Chen}}, \bibinfo {author} {\bibfnamefont {A.}~\bibnamefont {Verrier}},
  \bibinfo {author} {\bibfnamefont {D.}~\bibnamefont {Ziat}}, \bibinfo {author}
  {\bibfnamefont {A.~J.}\ \bibnamefont {Clune}}, \bibinfo {author}
  {\bibfnamefont {R.}~\bibnamefont {Rouane}}, \bibinfo {author} {\bibfnamefont
  {X.}~\bibnamefont {Bazier-Matte}}, \bibinfo {author} {\bibfnamefont
  {G.}~\bibnamefont {Wang}}, \bibinfo {author} {\bibfnamefont {S.}~\bibnamefont
  {Calder}}, \bibinfo {author} {\bibfnamefont {K.~M.}\ \bibnamefont {Taddei}},
  \bibinfo {author} {\bibfnamefont {C.~R.}\ \bibnamefont {dela Cruz}}, \bibinfo
  {author} {\bibfnamefont {A.~I.}\ \bibnamefont {Kolesnikov}}, \bibinfo
  {author} {\bibfnamefont {J.}~\bibnamefont {Ma}}, \bibinfo {author}
  {\bibfnamefont {J.-G.}\ \bibnamefont {Cheng}}, \bibinfo {author}
  {\bibfnamefont {Z.}~\bibnamefont {Liu}}, \bibinfo {author} {\bibfnamefont
  {J.~A.}\ \bibnamefont {Quilliam}}, \bibinfo {author} {\bibfnamefont {J.~L.}\
  \bibnamefont {Musfeldt}}, \bibinfo {author} {\bibfnamefont {H.~D.}\
  \bibnamefont {Zhou}},\ and\ \bibinfo {author} {\bibfnamefont {A.~A.}\
  \bibnamefont {Aczel}},\ }\bibfield  {title} {\bibinfo {title} {{Realization
  of the orbital-selective Mott state at the molecular level in
  ${\mathrm{Ba}}_{3}{\mathrm{LaRu}}_{2}{\mathrm{O}}_{9}$}},\ }\href
  {https://doi.org/10.1103/PhysRevMaterials.4.064409} {\bibfield  {journal}
  {\bibinfo  {journal} {Phys. Rev. Mater.}\ }\textbf {\bibinfo {volume} {4}},\
  \bibinfo {pages} {064409} (\bibinfo {year} {2020})}\BibitemShut {NoStop}%
\bibitem [{\citenamefont {Terasaki}\ \emph {et~al.}(2017)\citenamefont
  {Terasaki}, \citenamefont {Igarashi}, \citenamefont {Nagai}, \citenamefont
  {Tanabe}, \citenamefont {Taniguchi}, \citenamefont {Matsushita},
  \citenamefont {Wada}, \citenamefont {Takata}, \citenamefont {Kida},
  \citenamefont {Hagiwara}, \citenamefont {Kobayashi}, \citenamefont
  {Sagayama}, \citenamefont {Kumai}, \citenamefont {Nakao},\ and\ \citenamefont
  {Murakami}}]{Terasaki2017}%
  \BibitemOpen
  \bibfield  {author} {\bibinfo {author} {\bibfnamefont {I.}~\bibnamefont
  {Terasaki}}, \bibinfo {author} {\bibfnamefont {T.}~\bibnamefont {Igarashi}},
  \bibinfo {author} {\bibfnamefont {T.}~\bibnamefont {Nagai}}, \bibinfo
  {author} {\bibfnamefont {K.}~\bibnamefont {Tanabe}}, \bibinfo {author}
  {\bibfnamefont {H.}~\bibnamefont {Taniguchi}}, \bibinfo {author}
  {\bibfnamefont {T.}~\bibnamefont {Matsushita}}, \bibinfo {author}
  {\bibfnamefont {N.}~\bibnamefont {Wada}}, \bibinfo {author} {\bibfnamefont
  {A.}~\bibnamefont {Takata}}, \bibinfo {author} {\bibfnamefont
  {T.}~\bibnamefont {Kida}}, \bibinfo {author} {\bibfnamefont {M.}~\bibnamefont
  {Hagiwara}}, \bibinfo {author} {\bibfnamefont {K.}~\bibnamefont {Kobayashi}},
  \bibinfo {author} {\bibfnamefont {H.}~\bibnamefont {Sagayama}}, \bibinfo
  {author} {\bibfnamefont {R.}~\bibnamefont {Kumai}}, \bibinfo {author}
  {\bibfnamefont {H.}~\bibnamefont {Nakao}},\ and\ \bibinfo {author}
  {\bibfnamefont {Y.}~\bibnamefont {Murakami}},\ }\bibfield  {title} {\bibinfo
  {title} {{Absence of Magnetic Long Range Order in Ba$_{3}$ZnRu$_{2}$O$_{9}$:
  A Spin-Liquid Candidate in the $S = 3/2$ Dimer Lattice}},\ }\href
  {https://doi.org/10.7566/JPSJ.86.033702} {\bibfield  {journal} {\bibinfo
  {journal} {J. Phys. Soc. Jpn.}\ }\textbf {\bibinfo {volume} {86}},\ \bibinfo
  {pages} {033702} (\bibinfo {year} {2017})}\BibitemShut {NoStop}%
\bibitem [{\citenamefont {Yamamoto}\ \emph {et~al.}(2018)\citenamefont
  {Yamamoto}, \citenamefont {Taniguchi},\ and\ \citenamefont
  {Terasaki}}]{Yamamoto2018}%
  \BibitemOpen
  \bibfield  {author} {\bibinfo {author} {\bibfnamefont {T.~D.}\ \bibnamefont
  {Yamamoto}}, \bibinfo {author} {\bibfnamefont {H.}~\bibnamefont
  {Taniguchi}},\ and\ \bibinfo {author} {\bibfnamefont {I.}~\bibnamefont
  {Terasaki}},\ }\bibfield  {title} {\bibinfo {title} {{Dynamical coupling of
  dilute magnetic impurities with quantum spin liquid state in the $S=3/2$
  dimer compound Ba$_{3}$ZnRu$_{2}$O$_{9}$}},\ }\href
  {https://doi.org/10.1088/1361-648X/aad5ac} {\bibfield  {journal} {\bibinfo
  {journal} {J. Phys.: Condens. Matter}\ }\textbf {\bibinfo {volume} {30}},\
  \bibinfo {pages} {355801} (\bibinfo {year} {2018})}\BibitemShut {NoStop}%
\bibitem [{\citenamefont {Injac}\ \emph {et~al.}(2020)\citenamefont {Injac},
  \citenamefont {Solana-Madruga}, \citenamefont {Avdeev}, \citenamefont
  {Brand}, \citenamefont {Attfield},\ and\ \citenamefont
  {Kennedy}}]{Injac2020}%
  \BibitemOpen
  \bibfield  {author} {\bibinfo {author} {\bibfnamefont {S.}~\bibnamefont
  {Injac}}, \bibinfo {author} {\bibfnamefont {E.}~\bibnamefont
  {Solana-Madruga}}, \bibinfo {author} {\bibfnamefont {M.}~\bibnamefont
  {Avdeev}}, \bibinfo {author} {\bibfnamefont {H.~E.~A.}\ \bibnamefont
  {Brand}}, \bibinfo {author} {\bibfnamefont {J.~P.}\ \bibnamefont
  {Attfield}},\ and\ \bibinfo {author} {\bibfnamefont {B.~J.}\ \bibnamefont
  {Kennedy}},\ }\bibfield  {title} {\bibinfo {title} {{Studies of the 4d and 5d
  6H perovskites Ba$_3$BM$_2$O$_9${,} B = Ti{,} Zn{,} Y; M = Ru{,} Os{,} and
  cubic BaB$_{1/3}$Ru$_{2/3}$O$_3$ polymorphs stabilised under high
  pressure}},\ }\href {https://doi.org/10.1039/D0DT02349C} {\bibfield
  {journal} {\bibinfo  {journal} {Dalton Trans.}\ }\textbf {\bibinfo {volume}
  {49}},\ \bibinfo {pages} {12222} (\bibinfo {year} {2020})}\BibitemShut
  {NoStop}%
\bibitem [{\citenamefont {Sugano}\ \emph {et~al.}(1970)\citenamefont {Sugano},
  \citenamefont {Tanabe},\ and\ \citenamefont {Kamimura}}]{Sugano1970}%
  \BibitemOpen
  \bibfield  {author} {\bibinfo {author} {\bibfnamefont {S.}~\bibnamefont
  {Sugano}}, \bibinfo {author} {\bibfnamefont {Y.}~\bibnamefont {Tanabe}},\
  and\ \bibinfo {author} {\bibfnamefont {H.}~\bibnamefont {Kamimura}},\
  }\href@noop {} {\emph {\bibinfo {title} {{Transition-Metal Ions in
  Crystals}}}}\ (\bibinfo  {publisher} {Academic, New York},\ \bibinfo
  {year} {1970})\BibitemShut {NoStop}%
\bibitem [{\citenamefont {Tanaka}\ and\ \citenamefont
  {Hotta}(2020)}]{Tanaka2020}%
  \BibitemOpen
  \bibfield  {author} {\bibinfo {author} {\bibfnamefont {K.}~\bibnamefont
  {Tanaka}}\ and\ \bibinfo {author} {\bibfnamefont {C.}~\bibnamefont {Hotta}},\
  }\bibfield  {title} {\bibinfo {title} {{Multiple quadrupolar or nematic
  phases driven by the Heisenberg interactions in a spin-1 dimer system forming
  a bilayer}},\ }\href {https://doi.org/10.1103/PhysRevB.101.094422} {\bibfield
   {journal} {\bibinfo  {journal} {Phys. Rev. B}\ }\textbf {\bibinfo {volume}
  {101}},\ \bibinfo {pages} {094422} (\bibinfo {year} {2020})}\BibitemShut
  {NoStop}%
\bibitem [{\citenamefont {Streltsov}\ and\ \citenamefont
  {Khomskii}(2017)}]{Streltsov2017}%
  \BibitemOpen
  \bibfield  {author} {\bibinfo {author} {\bibfnamefont {S.~V.}\ \bibnamefont
  {Streltsov}}\ and\ \bibinfo {author} {\bibfnamefont {D.~I.}\ \bibnamefont
  {Khomskii}},\ }\bibfield  {title} {\bibinfo {title} {{Orbital physics in
  transition metal compounds: new trends}},\ }\href
  {https://doi.org/10.3367/UFNe.2017.08.038196} {\bibfield  {journal} {\bibinfo
   {journal} {Physics-Uspekhi}\ }\textbf {\bibinfo {volume} {60}},\ \bibinfo
  {pages} {1121} (\bibinfo {year} {2017})}\BibitemShut {NoStop}%
\bibitem [{\citenamefont {Hong}\ and\ \citenamefont
  {Sleight}(1997)}]{Hong1997}%
  \BibitemOpen
  \bibfield  {author} {\bibinfo {author} {\bibfnamefont {S.-T.}\ \bibnamefont
  {Hong}}\ and\ \bibinfo {author} {\bibfnamefont {A.~W.}\ \bibnamefont
  {Sleight}},\ }\bibfield  {title} {\bibinfo {title} {{Crystal Structure of 4H
  BaRuO$_3$: High Pressure Phase Prepared at Ambient Pressure}},\ }\href
  {https://doi.org/https://doi.org/10.1006/jssc.1996.7200} {\bibfield
  {journal} {\bibinfo  {journal} {J. Solid State Chem.}\ }\textbf {\bibinfo
  {volume} {128}},\ \bibinfo {pages} {251} (\bibinfo {year}
  {1997})}\BibitemShut {NoStop}%
\bibitem [{\citenamefont {Gretarsson}\ \emph {et~al.}(2020)\citenamefont
  {Gretarsson}, \citenamefont {Ketenoglu}, \citenamefont {Harder},
  \citenamefont {Mayer}, \citenamefont {Dill}, \citenamefont {Spiwek},
  \citenamefont {Schulte-Schrepping}, \citenamefont {Tischer}, \citenamefont
  {Wille}, \citenamefont {Keimer},\ and\ \citenamefont
  {Yava{\c{s}}}}]{Gretarsson2020}%
  \BibitemOpen
  \bibfield  {author} {\bibinfo {author} {\bibfnamefont {H.}~\bibnamefont
  {Gretarsson}}, \bibinfo {author} {\bibfnamefont {D.}~\bibnamefont
  {Ketenoglu}}, \bibinfo {author} {\bibfnamefont {M.}~\bibnamefont {Harder}},
  \bibinfo {author} {\bibfnamefont {S.}~\bibnamefont {Mayer}}, \bibinfo
  {author} {\bibfnamefont {F.-U.}\ \bibnamefont {Dill}}, \bibinfo {author}
  {\bibfnamefont {M.}~\bibnamefont {Spiwek}}, \bibinfo {author} {\bibfnamefont
  {H.}~\bibnamefont {Schulte-Schrepping}}, \bibinfo {author} {\bibfnamefont
  {M.}~\bibnamefont {Tischer}}, \bibinfo {author} {\bibfnamefont {H.-C.}\
  \bibnamefont {Wille}}, \bibinfo {author} {\bibfnamefont {B.}~\bibnamefont
  {Keimer}},\ and\ \bibinfo {author} {\bibfnamefont {H.}~\bibnamefont
  {Yava{\c{s}}}},\ }\bibfield  {title} {\bibinfo {title} {{IRIXS: a resonant
  inelastic X-ray scattering instrument dedicated to X-rays in the intermediate
  energy range}},\ }\href {https://doi.org/10.1107/S1600577519017119}
  {\bibfield  {journal} {\bibinfo  {journal} {J. Synch. Rad.}\ }\textbf
  {\bibinfo {volume} {27}},\ \bibinfo {pages} {538} (\bibinfo {year}
  {2020})}\BibitemShut {NoStop}%
\bibitem [{\citenamefont {Gretarsson}\ \emph {et~al.}(2019)\citenamefont
  {Gretarsson}, \citenamefont {Suzuki}, \citenamefont {Kim}, \citenamefont
  {Ueda}, \citenamefont {Krautloher}, \citenamefont {Kim}, \citenamefont
  {Yava\ifmmode~\mbox{\c{s}}\else \c{s}\fi{}}, \citenamefont {Khaliullin},\
  and\ \citenamefont {Keimer}}]{Gretarsson2019}%
  \BibitemOpen
  \bibfield  {author} {\bibinfo {author} {\bibfnamefont {H.}~\bibnamefont
  {Gretarsson}}, \bibinfo {author} {\bibfnamefont {H.}~\bibnamefont {Suzuki}},
  \bibinfo {author} {\bibfnamefont {H.}~\bibnamefont {Kim}}, \bibinfo {author}
  {\bibfnamefont {K.}~\bibnamefont {Ueda}}, \bibinfo {author} {\bibfnamefont
  {M.}~\bibnamefont {Krautloher}}, \bibinfo {author} {\bibfnamefont {B.~J.}\
  \bibnamefont {Kim}}, \bibinfo {author} {\bibfnamefont {H.}~\bibnamefont
  {Yava\ifmmode~\mbox{\c{s}}\else \c{s}\fi{}}}, \bibinfo {author}
  {\bibfnamefont {G.}~\bibnamefont {Khaliullin}},\ and\ \bibinfo {author}
  {\bibfnamefont {B.}~\bibnamefont {Keimer}},\ }\bibfield  {title} {\bibinfo
  {title} {{Observation of spin-orbit excitations and Hund's multiplets in
  ${\mathrm{Ca}}_{2}{\mathrm{RuO}}_{4}$}},\ }\href
  {https://doi.org/10.1103/PhysRevB.100.045123} {\bibfield  {journal} {\bibinfo
   {journal} {Phys. Rev. B}\ }\textbf {\bibinfo {volume} {100}},\ \bibinfo
  {pages} {045123} (\bibinfo {year} {2019})}\BibitemShut {NoStop}%
\bibitem [{\citenamefont {Bertinshaw}\ \emph {et~al.}(2021)\citenamefont
  {Bertinshaw}, \citenamefont {Krautloher}, \citenamefont {Suzuki},
  \citenamefont {Takahashi}, \citenamefont {Ivanov}, \citenamefont
  {Yava\ifmmode~\mbox{\c{s}}\else \c{s}\fi{}}, \citenamefont {Kim},
  \citenamefont {Gretarsson},\ and\ \citenamefont {Keimer}}]{Bertinshaw2021}%
  \BibitemOpen
  \bibfield  {author} {\bibinfo {author} {\bibfnamefont {J.}~\bibnamefont
  {Bertinshaw}}, \bibinfo {author} {\bibfnamefont {M.}~\bibnamefont
  {Krautloher}}, \bibinfo {author} {\bibfnamefont {H.}~\bibnamefont {Suzuki}},
  \bibinfo {author} {\bibfnamefont {H.}~\bibnamefont {Takahashi}}, \bibinfo
  {author} {\bibfnamefont {A.}~\bibnamefont {Ivanov}}, \bibinfo {author}
  {\bibfnamefont {H.}~\bibnamefont {Yava\ifmmode~\mbox{\c{s}}\else
  \c{s}\fi{}}}, \bibinfo {author} {\bibfnamefont {B.~J.}\ \bibnamefont {Kim}},
  \bibinfo {author} {\bibfnamefont {H.}~\bibnamefont {Gretarsson}},\ and\
  \bibinfo {author} {\bibfnamefont {B.}~\bibnamefont {Keimer}},\ }\bibfield
  {title} {\bibinfo {title} {{Spin and charge excitations in the correlated
  multiband metal ${\mathrm{Ca}}_{3}{\mathrm{Ru}}_{2}{\mathrm{O}}_{7}$}},\
  }\href {https://doi.org/10.1103/PhysRevB.103.085108} {\bibfield  {journal}
  {\bibinfo  {journal} {Phys. Rev. B}\ }\textbf {\bibinfo {volume} {103}},\
  \bibinfo {pages} {085108} (\bibinfo {year} {2021})}\BibitemShut {NoStop}%
\bibitem [{\citenamefont {Suzuki}\ \emph {et~al.}(2023)\citenamefont {Suzuki},
  \citenamefont {Wang}, \citenamefont {Bertinshaw}, \citenamefont {Strand},
  \citenamefont {K{\"a}ser}, \citenamefont {Krautloher}, \citenamefont {Yang},
  \citenamefont {Wentzell}, \citenamefont {Parcollet}, \citenamefont
  {Jerzembeck}, \citenamefont {Kikugawa}, \citenamefont {Mackenzie},
  \citenamefont {Georges}, \citenamefont {Hansmann}, \citenamefont
  {Gretarsson},\ and\ \citenamefont {Keimer}}]{Suzuki2023}%
  \BibitemOpen
  \bibfield  {author} {\bibinfo {author} {\bibfnamefont {H.}~\bibnamefont
  {Suzuki}}, \bibinfo {author} {\bibfnamefont {L.}~\bibnamefont {Wang}},
  \bibinfo {author} {\bibfnamefont {J.}~\bibnamefont {Bertinshaw}}, \bibinfo
  {author} {\bibfnamefont {H.~U.~R.}\ \bibnamefont {Strand}}, \bibinfo {author}
  {\bibfnamefont {S.}~\bibnamefont {K{\"a}ser}}, \bibinfo {author}
  {\bibfnamefont {M.}~\bibnamefont {Krautloher}}, \bibinfo {author}
  {\bibfnamefont {Z.}~\bibnamefont {Yang}}, \bibinfo {author} {\bibfnamefont
  {N.}~\bibnamefont {Wentzell}}, \bibinfo {author} {\bibfnamefont
  {O.}~\bibnamefont {Parcollet}}, \bibinfo {author} {\bibfnamefont
  {F.}~\bibnamefont {Jerzembeck}}, \bibinfo {author} {\bibfnamefont
  {N.}~\bibnamefont {Kikugawa}}, \bibinfo {author} {\bibfnamefont {A.~P.}\
  \bibnamefont {Mackenzie}}, \bibinfo {author} {\bibfnamefont {A.}~\bibnamefont
  {Georges}}, \bibinfo {author} {\bibfnamefont {P.}~\bibnamefont {Hansmann}},
  \bibinfo {author} {\bibfnamefont {H.}~\bibnamefont {Gretarsson}},\ and\
  \bibinfo {author} {\bibfnamefont {B.}~\bibnamefont {Keimer}},\ }\bibfield
  {title} {\bibinfo {title} {{Distinct spin and orbital dynamics in
  Sr$_2$RuO$_4$}},\ }\href {https://doi.org/10.1038/s41467-023-42804-3}
  {\bibfield  {journal} {\bibinfo  {journal} {Nat. Commun.}\ }\textbf {\bibinfo
  {volume} {14}},\ \bibinfo {pages} {7042} (\bibinfo {year}
  {2023})}\BibitemShut {NoStop}%
\bibitem [{\citenamefont {Zhao}\ \emph {et~al.}(2007)\citenamefont {Zhao},
  \citenamefont {Yang}, \citenamefont {Yu}, \citenamefont {Li}, \citenamefont
  {Yu}, \citenamefont {Fang}, \citenamefont {Chen},\ and\ \citenamefont
  {Jin}}]{Zhao2007}%
  \BibitemOpen
  \bibfield  {author} {\bibinfo {author} {\bibfnamefont {J.}~\bibnamefont
  {Zhao}}, \bibinfo {author} {\bibfnamefont {L.}~\bibnamefont {Yang}}, \bibinfo
  {author} {\bibfnamefont {Y.}~\bibnamefont {Yu}}, \bibinfo {author}
  {\bibfnamefont {F.}~\bibnamefont {Li}}, \bibinfo {author} {\bibfnamefont
  {R.}~\bibnamefont {Yu}}, \bibinfo {author} {\bibfnamefont {Z.}~\bibnamefont
  {Fang}}, \bibinfo {author} {\bibfnamefont {L.}~\bibnamefont {Chen}},\ and\
  \bibinfo {author} {\bibfnamefont {C.}~\bibnamefont {Jin}},\ }\bibfield
  {title} {\bibinfo {title} {{Structural and physical properties of the 6H
  BaRuO$_{3}$ polymorph synthesized under high pressure}},\ }\href
  {https://doi.org/https://doi.org/10.1016/j.jssc.2007.07.031} {\bibfield
  {journal} {\bibinfo  {journal} {J. Solid State Chem.}\ }\textbf {\bibinfo
  {volume} {180}},\ \bibinfo {pages} {2816} (\bibinfo {year}
  {2007})}\BibitemShut {NoStop}%
\bibitem [{\citenamefont {Suzuki}\ \emph {et~al.}(2019)\citenamefont {Suzuki},
  \citenamefont {Gretarsson}, \citenamefont {Ishikawa}, \citenamefont {Ueda},
  \citenamefont {Yang}, \citenamefont {Liu}, \citenamefont {Kim}, \citenamefont
  {Kukusta}, \citenamefont {Yaresko}, \citenamefont {Minola}, \citenamefont
  {Sears}, \citenamefont {Francoual}, \citenamefont {Wille}, \citenamefont
  {Nuss}, \citenamefont {Takagi}, \citenamefont {Kim}, \citenamefont
  {Khaliullin}, \citenamefont {Yava{\c{s}}},\ and\ \citenamefont
  {Keimer}}]{Suzuki2019}%
  \BibitemOpen
  \bibfield  {author} {\bibinfo {author} {\bibfnamefont {H.}~\bibnamefont
  {Suzuki}}, \bibinfo {author} {\bibfnamefont {H.}~\bibnamefont {Gretarsson}},
  \bibinfo {author} {\bibfnamefont {H.}~\bibnamefont {Ishikawa}}, \bibinfo
  {author} {\bibfnamefont {K.}~\bibnamefont {Ueda}}, \bibinfo {author}
  {\bibfnamefont {Z.}~\bibnamefont {Yang}}, \bibinfo {author} {\bibfnamefont
  {H.}~\bibnamefont {Liu}}, \bibinfo {author} {\bibfnamefont {H.}~\bibnamefont
  {Kim}}, \bibinfo {author} {\bibfnamefont {D.}~\bibnamefont {Kukusta}},
  \bibinfo {author} {\bibfnamefont {A.}~\bibnamefont {Yaresko}}, \bibinfo
  {author} {\bibfnamefont {M.}~\bibnamefont {Minola}}, \bibinfo {author}
  {\bibfnamefont {J.~A.}\ \bibnamefont {Sears}}, \bibinfo {author}
  {\bibfnamefont {S.}~\bibnamefont {Francoual}}, \bibinfo {author}
  {\bibfnamefont {H.-C.}\ \bibnamefont {Wille}}, \bibinfo {author}
  {\bibfnamefont {J.}~\bibnamefont {Nuss}}, \bibinfo {author} {\bibfnamefont
  {H.}~\bibnamefont {Takagi}}, \bibinfo {author} {\bibfnamefont {B.~J.}\
  \bibnamefont {Kim}}, \bibinfo {author} {\bibfnamefont {G.}~\bibnamefont
  {Khaliullin}}, \bibinfo {author} {\bibfnamefont {H.}~\bibnamefont
  {Yava{\c{s}}}},\ and\ \bibinfo {author} {\bibfnamefont {B.}~\bibnamefont
  {Keimer}},\ }\bibfield  {title} {\bibinfo {title} {{Spin waves and spin-state
  transitions in a ruthenate high-temperature antiferromagnet}},\ }\href
  {https://doi.org/10.1038/s41563-019-0327-2} {\bibfield  {journal} {\bibinfo
  {journal} {Nat. Mater.}\ }\textbf {\bibinfo {volume} {18}},\ \bibinfo {pages}
  {563} (\bibinfo {year} {2019})}\BibitemShut {NoStop}%
\bibitem [{\citenamefont {Braicovich}\ \emph {et~al.}(2020)\citenamefont
  {Braicovich}, \citenamefont {Rossi}, \citenamefont {Fumagalli}, \citenamefont
  {Peng}, \citenamefont {Wang}, \citenamefont {Arpaia}, \citenamefont {Betto},
  \citenamefont {De~Luca}, \citenamefont {Di~Castro}, \citenamefont {Kummer},
  \citenamefont {Moretti~Sala}, \citenamefont {Pagetti}, \citenamefont
  {Balestrino}, \citenamefont {Brookes}, \citenamefont {Salluzzo},
  \citenamefont {Johnston}, \citenamefont {van~den Brink},\ and\ \citenamefont
  {Ghiringhelli}}]{Braicovich2020}%
  \BibitemOpen
  \bibfield  {author} {\bibinfo {author} {\bibfnamefont {L.}~\bibnamefont
  {Braicovich}}, \bibinfo {author} {\bibfnamefont {M.}~\bibnamefont {Rossi}},
  \bibinfo {author} {\bibfnamefont {R.}~\bibnamefont {Fumagalli}}, \bibinfo
  {author} {\bibfnamefont {Y.}~\bibnamefont {Peng}}, \bibinfo {author}
  {\bibfnamefont {Y.}~\bibnamefont {Wang}}, \bibinfo {author} {\bibfnamefont
  {R.}~\bibnamefont {Arpaia}}, \bibinfo {author} {\bibfnamefont
  {D.}~\bibnamefont {Betto}}, \bibinfo {author} {\bibfnamefont {G.~M.}\
  \bibnamefont {De~Luca}}, \bibinfo {author} {\bibfnamefont {D.}~\bibnamefont
  {Di~Castro}}, \bibinfo {author} {\bibfnamefont {K.}~\bibnamefont {Kummer}},
  \bibinfo {author} {\bibfnamefont {M.}~\bibnamefont {Moretti~Sala}}, \bibinfo
  {author} {\bibfnamefont {M.}~\bibnamefont {Pagetti}}, \bibinfo {author}
  {\bibfnamefont {G.}~\bibnamefont {Balestrino}}, \bibinfo {author}
  {\bibfnamefont {N.~B.}\ \bibnamefont {Brookes}}, \bibinfo {author}
  {\bibfnamefont {M.}~\bibnamefont {Salluzzo}}, \bibinfo {author}
  {\bibfnamefont {S.}~\bibnamefont {Johnston}}, \bibinfo {author}
  {\bibfnamefont {J.}~\bibnamefont {van~den Brink}},\ and\ \bibinfo {author}
  {\bibfnamefont {G.}~\bibnamefont {Ghiringhelli}},\ }\bibfield  {title}
  {\bibinfo {title} {{Determining the electron-phonon coupling in
  superconducting cuprates by resonant inelastic x-ray scattering: Methods and
  results on
  ${\mathrm{Nd}}_{1+x}{\mathrm{Ba}}_{2\ensuremath{-}x}{\mathrm{Cu}}_{3}{\mathrm{O}}_{7\ensuremath{-}\ensuremath{\delta}}$}},\
  }\href {https://doi.org/10.1103/PhysRevResearch.2.023231} {\bibfield
  {journal} {\bibinfo  {journal} {Phys. Rev. Res.}\ }\textbf {\bibinfo {volume}
  {2}},\ \bibinfo {pages} {023231} (\bibinfo {year} {2020})}\BibitemShut
  {NoStop}%
\bibitem [{\citenamefont {Zapf}\ \emph {et~al.}(2014)\citenamefont {Zapf},
  \citenamefont {Jaime},\ and\ \citenamefont {Batista}}]{Zapf2014}%
  \BibitemOpen
  \bibfield  {author} {\bibinfo {author} {\bibfnamefont {V.}~\bibnamefont
  {Zapf}}, \bibinfo {author} {\bibfnamefont {M.}~\bibnamefont {Jaime}},\ and\
  \bibinfo {author} {\bibfnamefont {C.~D.}\ \bibnamefont {Batista}},\
  }\bibfield  {title} {\bibinfo {title} {{Bose-Einstein condensation in quantum
  magnets}},\ }\href {https://doi.org/10.1103/RevModPhys.86.563} {\bibfield
  {journal} {\bibinfo  {journal} {Rev. Mod. Phys.}\ }\textbf {\bibinfo {volume}
  {86}},\ \bibinfo {pages} {563} (\bibinfo {year} {2014})}\BibitemShut
  {NoStop}%
\bibitem [{\citenamefont {Hayashida}\ \emph {et~al.}(2024)\citenamefont
  {Hayashida}, \citenamefont {Gretarsson}, \citenamefont {Puphal},
  \citenamefont {Isobe}, \citenamefont {Goering}, \citenamefont {Matsumoto},
  \citenamefont {Nuss}, \citenamefont {Takagi}, \citenamefont {Hepting},\ and\
  \citenamefont {Keimer}}]{dataset}%
  \BibitemOpen
  \bibfield  {author} {\bibinfo {author} {\bibfnamefont {S.}~\bibnamefont
  {Hayashida}}, \bibinfo {author} {\bibfnamefont {H.}~\bibnamefont
  {Gretarsson}}, \bibinfo {author} {\bibfnamefont {P.}~\bibnamefont {Puphal}},
  \bibinfo {author} {\bibfnamefont {M.}~\bibnamefont {Isobe}}, \bibinfo
  {author} {\bibfnamefont {E.}~\bibnamefont {Goering}}, \bibinfo {author}
  {\bibfnamefont {Y.}~\bibnamefont {Matsumoto}}, \bibinfo {author}
  {\bibfnamefont {J.}~\bibnamefont {Nuss}}, \bibinfo {author} {\bibfnamefont
  {H.}~\bibnamefont {Takagi}}, \bibinfo {author} {\bibfnamefont
  {M.}~\bibnamefont {Hepting}},\ and\ \bibinfo {author} {\bibfnamefont
  {B.}~\bibnamefont {Keimer}},\ }\bibfield  {title} {\bibinfo {title}
  {{Magnetic ground state of the dimer-based hexagonal perovskite
  Ba$_{3}$ZnRu$_{2}$O$_{9}$ (data set)}},\ }\href
  {https://doi.org/10.5281/zenodo.14969237} {10.5281/zenodo.14969237} (\bibinfo
  {year} {2024})\BibitemShut {NoStop}%
\bibitem [{\citenamefont {Sheldrick}(2008)}]{Sheldrick2008}%
  \BibitemOpen
  \bibfield  {author} {\bibinfo {author} {\bibfnamefont {G.~M.}\ \bibnamefont
  {Sheldrick}},\ }\bibfield  {title} {\bibinfo {title} {{A short history of
  {\it SHELX}}},\ }\href {https://doi.org/10.1107/S0108767307043930} {\bibfield
   {journal} {\bibinfo  {journal} {Acta Crystallogr. A}\ }\textbf {\bibinfo
  {volume} {64}},\ \bibinfo {pages} {112} (\bibinfo {year} {2008})}\BibitemShut
  {NoStop}%
\bibitem [{\citenamefont {Sheldrick}(2015)}]{Sheldrick2014}%
  \BibitemOpen
  \bibfield  {author} {\bibinfo {author} {\bibfnamefont {G.~M.}\ \bibnamefont
  {Sheldrick}},\ }\bibfield  {title} {\bibinfo {title} {{Crystal structure
  refinement with {\it SHELXL}}},\ }\href
  {https://doi.org/10.1107/S2053229614024218} {\bibfield  {journal} {\bibinfo
  {journal} {Acta Crystallogr. C}\ }\textbf {\bibinfo {volume} {71}},\ \bibinfo
  {pages} {3} (\bibinfo {year} {2015})}\BibitemShut {NoStop}%
\bibitem [{\citenamefont {Rodr{\'i}guez-Carvajal}(1993)}]{FullProf}%
  \BibitemOpen
  \bibfield  {author} {\bibinfo {author} {\bibfnamefont {J.}~\bibnamefont
  {Rodr{\'i}guez-Carvajal}},\ }\bibfield  {title} {\bibinfo {title} {{Recent
  advances in magnetic structure determination by neutron powder
  diffraction}},\ }\href
  {https://www.sciencedirect.com/science/article/pii/092145269390108I}
  {\bibfield  {journal} {\bibinfo  {journal} {Physica B: Condensed Matter}\
  }\textbf {\bibinfo {volume} {192}},\ \bibinfo {pages} {55} (\bibinfo {year}
  {1993})}\BibitemShut {NoStop}%
\end{thebibliography}
%

\end{document}